\documentclass[format=sigconf, review=false]{acmart}
\usepackage[utf8]{inputenc}
\usepackage{amsmath}
\usepackage{listings}
\usepackage{cleveref}
\usepackage{tikz}
\usepackage{booktabs}
\usepackage{enumitem}
\usepackage{pifont}
\usepackage{xspace}
\usepackage{wrapfig}

\newcommand{\one}{\ding{192}\xspace}
\newcommand{\two}{\ding{193}\xspace}
\newcommand{\three}{\ding{194}\xspace}

\usetikzlibrary{positioning,arrows,calc,shapes,shapes.geometric,shapes.misc}
\tikzset{
    on grid,
    auto,
    ip/.style = {
        draw,
        fill=white,
        shape=rectangle,
        minimum height=3em,
        minimum width=3em,
        line width=1pt
    },
    core/.style = {
        draw,
        fill=white,
        shape=rectangle,
        minimum height=3em,
        minimum width=3em,
        line width=1pt,
        dashed
    },
    map_entry/.style = {
        trapezium,
        trapezium angle=60,
        draw,
        fill=gray!10
    },
    map_exit/.style = {
        trapezium,
        trapezium angle=-60,
        draw,
        fill=gray!10
    },
    array/.style = {
        ellipse,
        draw,
        align=center,
        fill=gray!10
    },
    stream/.style = {
        ellipse,
        draw,
        align=center,
        fill=gray!10,
        dashed
    },
    tasklet/.style = {
        draw,
        chamfered rectangle,
        fill=gray!10
    }
}

\title{Temporal Vectorization: A Compiler Approach to Automatic Multi-Pumping}

\author{Carl-Johannes Johnsen}
\affiliation{
    \institution{University of Copenhagen}
    \department{Department of Computer Science}
    \streetaddress{Universitetsparken 5}
    \city{Copenhagen}
    \country{Denmark}
    \postcode{2100}
}
\email{carl-johannes@di.ku.dk}

\author{Tiziano De Matteis}
\affiliation{
    \institution{ETH Zurich}
    \department{Department of Computer Science}
    \city{Zurich}
    \country{Switzerland}
}
\email{tdematt@inf.ethz.ch}

\author{Tal Ben-Nun}
\affiliation{
    \institution{ETH Zurich}
    \department{Department of Computer Science}
    \city{Zurich}
    \country{Switzerland}
}
\email{talbn@inf.ethz.ch}

\author{Johannes de Fine Licht}
\affiliation{
    \institution{ETH Zurich}
    \department{Department of Computer Science}
    \city{Zurich}
    \country{Switzerland}
}
\email{definelicht@inf.ethz.ch}

\author{Torsten Hoefler}
\affiliation{
    \institution{ETH Zurich}
    \department{Department of Computer Science}
    \city{Zurich}
    \country{Switzerland}
}
\email{htor@inf.ethz.ch}

\begin{document}
\settopmatter{printacmref=false}
\setcopyright{none}
\renewcommand\footnotetextcopyrightpermission[1]{}
\pagestyle{plain}

\setcopyright{none}
\makeatletter
\def\@copyrightspace{\relax}
\makeatother

\begin{abstract}
    The multi-pumping resource sharing technique can overcome the limitations commonly found in single-clocked FPGA designs by allowing hardware components to operate at a higher clock frequency than the surrounding system. However, this optimization cannot be expressed in high levels of abstraction, such as HLS, requiring the use of hand-optimized RTL. In this paper we show how to leverage multiple clock domains for computational subdomains on reconfigurable devices through data movement analysis on high-level programs.
    We offer a novel view on multi-pumping as a compiler optimization --- a superclass of traditional vectorization.
    As multiple data elements are fed and consumed, the computations are packed temporally rather than spatially.
    The optimization is applied automatically using an intermediate representation that maps high-level code to HLS. Internally, the optimization injects modules into the generated designs, incorporating RTL for fine-grained control over the clock domains.
    We obtain a reduction of resource consumption by up to 50\% on critical components and 23\% on average. For scalable designs, this can enable further parallelism, increasing overall performance.
\end{abstract}

\maketitle

\section{Introduction}

Designing application-specific hardware is fundamentally resource-constrained.
The performance of a circuit implementing a parallel computation is the product of the degree of parallelism and the frequency at which it is clocked.
We can improve the performance by either increasing the circuit's frequency, or by introducing additional parallelism if opportunities exist in the application.
Viewed differently, if we fix the performance of a circuit, we can reduce the resource usage by proportionally increasing the frequency.
This resource reduction can either be used to increase the parallelism of the computation, or to implement other operations.

In addition to power and thermal limits, increasing the frequency of a design is not always possible due to timing constraints.
However, not all subdomains need to run at the same frequency.
Some subdomains might not be critical for performance or contribute significantly to overall resource utilization, and thus do not bottleneck throughput at lower frequencies.
For subdomains that are primarily concerned with moving data, such as I/O interfaces accessing DRAM, PCIe, or networking, we can get away with increasing the width of the data path rather than increasing the frequency.

We can thus increase the frequency of only the resource and/or performance-critical subdomains, feeding more data per cycle from the slower clock domain. This is referred to as ``multi-pumping''~\cite{multi-pumping}.
\begin{figure}
    \centering
    \includegraphics[width=\columnwidth]{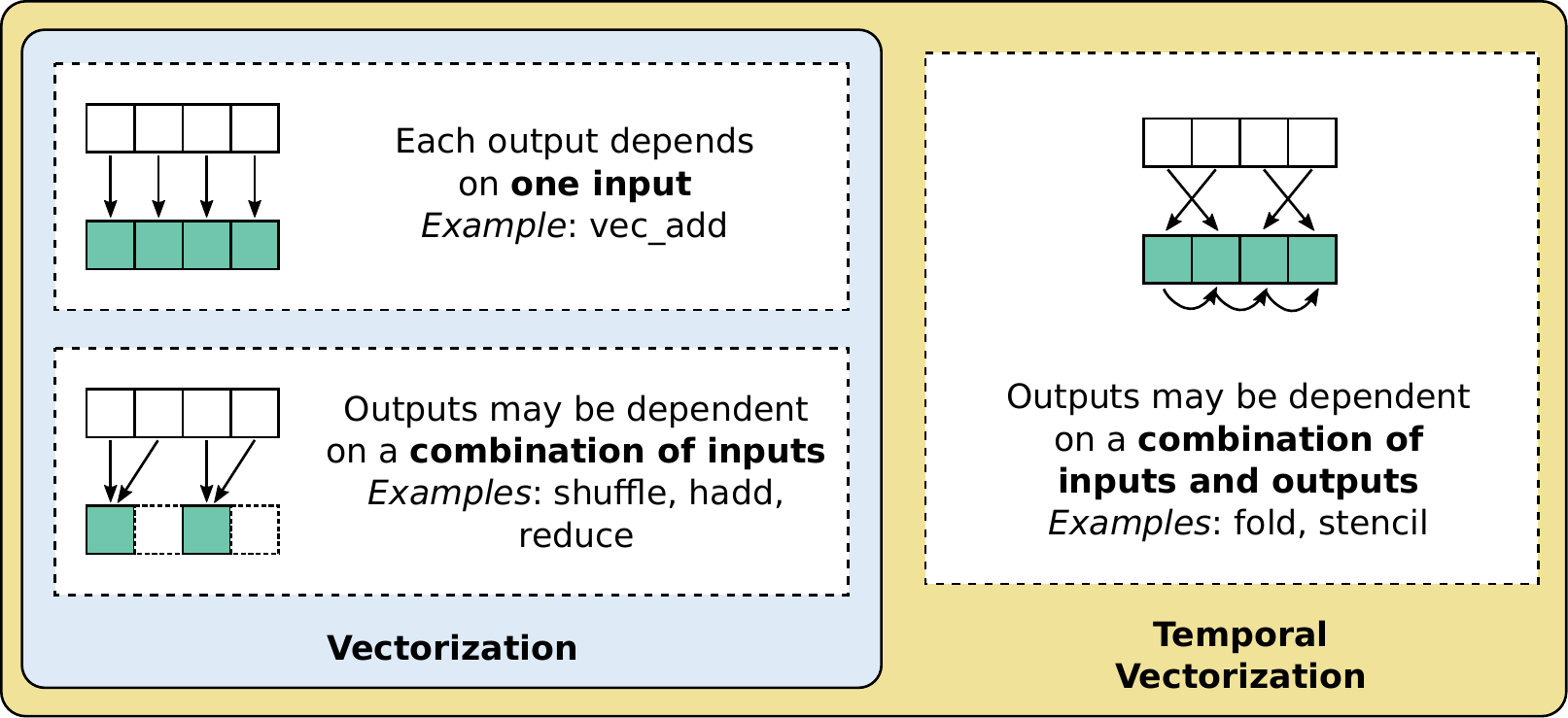}
    \vspace{-2em}
    \caption{Temporal Vectorization}
    \vspace{-2em}

    \label{fig:temporal_vectorization}
\end{figure}
Computational components are usually densely connected by \emph{short} paths, while the data paths connecting them account for \emph{long} data paths across the chip.
From a data movement perspective, multi-pumping can be seen as a form of ``temporal vectorization'': the \emph{long} data paths leading to and/or from the computation are widened, but the densely connected logic performing the computation itself is left unchanged, while conceptually being ``vectorized'' across multiple clock cycles.
As a result, the densely connected short paths now only have to meet timing at a higher frequency locally, while the long data paths do not need additional buffering of the signal that would otherwise be required at higher frequencies.

Vectorization is traditionally applied on computations where each output depends only on one input (such as in the case of vector addition), or when the output depends on a combination of inputs (such as, in horizontal addition or reductions). Temporal vectorization relaxes these requirements:
it must still be possible to parallelize the memory source/destination, but it does not impose any requirements on the computation -- the computation does not even need to be analyzable, and dependencies between operations are allowed without any additional handling. For this reason, temporal vectorization can bee seen as a \emph{superclass} of traditional vectorization (see \Cref{fig:temporal_vectorization}).

Introducing multi-pumping to a design is an invasive procedure that requires significant effort, requiring clock domain crossing and data width conversion to be performed at either end of the higher clocked domain.
In high-level synthesis (HLS) development flows in particular, multi-pumping is either not supported altogether or severely limited in scope, resulting in this optimization rarely being exploited for FPGA development in practice.

In this work, we show how the multi-pumping optimization can be automatically applied using data movement analysis.
By capturing all data movement to and from computational subdomains, we can identify if they can be multi-pumped, introduce the new clock domain, and insert the necessary domain crossing logic.
We demonstrate automatic multi-pumping on several applications compiled from a Python frontend implementation to FPGA architectures on a Xilinx accelerator board.
We show how resource usage of critical FPGA resources, such as DSPs and BRAM, is reduced by 50\% when critical subdomains are double-pumped, and how this can be used in relevant applications to increase the overall performance of the design by exploiting the resources freed up by multi-pumping.

In particular, the main contributions of this paper are:
\begin{itemize}
    \item A novel view of the multi-pumping optimization as temporal vectorization.
    \item Automatic application of the multi-pumping optimization to the broader scope of computational subdomains, rather than individual components.
    \item The ability for software developers to exploit the multi-pumping optimization in high-level code by providing automatic HLS and RTL integration.
    \item Demonstrating the benefits of multi-pumping optimization in performance increase or resource reduction on four different use cases.
\end{itemize}

\section{Multi-pumping}
Programming FPGAs with HLS revolves around designing deep hardware pipelines, exploiting the spatial parallelism offered by the device. Optimizing compilers and performance engineers leverage classical high-performance computing and FPGA-oriented transformations to achieve this goal~\cite{hls_transformations}. Resource utilization is a metric that must be considered in optimizing code for FPGA, as space consumption can be one of the critical factors limiting the performance of FPGA large-scale designs.

Traditionally, resource-sharing techniques have been used to reduce area consumption, but these usually come at the expense of degraded circuit performance.  Multi-pumping aims to overcome the limitations of other solutions by exploiting the capability of the hardware fabric to run different components at different clock rates~\cite{multi-pumping,  13-multipump, 16-multidsp}. FPGA designs created with modern HLS tools typically run at 200-350 Mhz, while other FPGA components, such as DSPs or on-chip memory, can be clocked at a higher frequency.
For example, the DSP48 block of a Xilinx Alveo U280 can be clocked up to $891$ MHz~\cite{xilinx-fmax}, almost three times higher than the usual design frequency achieved by HLS. While reaching such higher frequencies is infeasible (due to routing and timing closure requirements), it is clear that internal components are not fully exploited in high-level FPGA designs.

\subsection{Exploiting multiple clock domains}\label{sec:exploiting-clock-domains}

FPGA designs usually have a single clock domain, where the entire design shares the same clock signal.
To apply multi-pumping, we need to have two clock domains, one for the slowly clocked components, such as reader/writer to external memory, and one highly clocked for the internal compute components.

\begin{figure}[t]
    \centering
    \includegraphics[width=\linewidth]{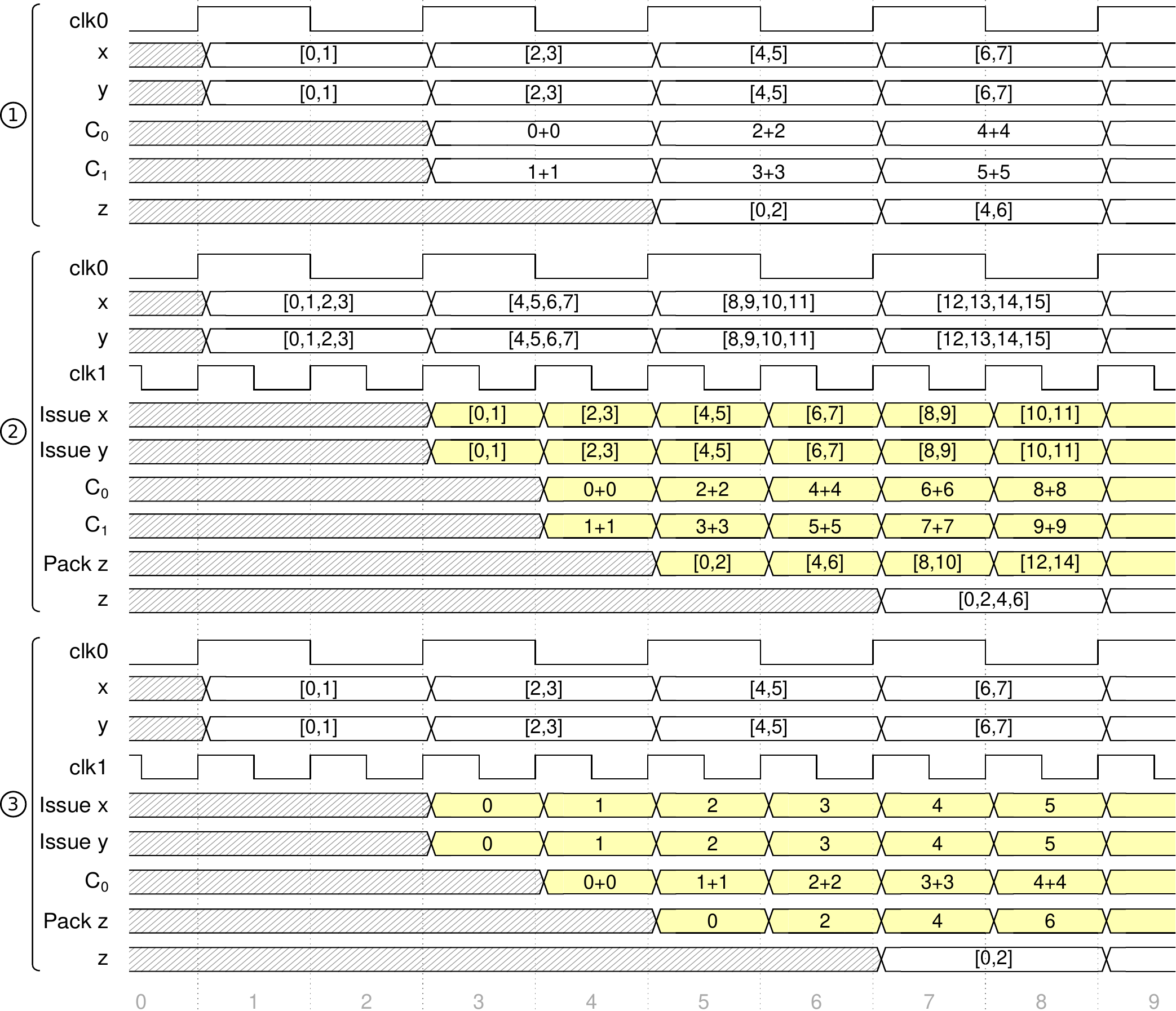}
    % Generated through https://wavedrom.com/
    \caption{Waveforms depicting original implementation and multi-pumping approaches for vector addition, with $M=2, v=2$.}

    \label{fig:waveform}
    \vspace{-3mm}
\end{figure}

Consider the case of a $V$-way vectorized vector addition $z=x+y$, where $V$ elements of $x$ and $y$ are read every tick of the clock $clk_0$. To process the entire vector, the internal components $C$, adding together a single element of $x$ and $y$, have to be replicated $V$ times. Picture \one in  \Cref{fig:waveform} shows a waveform describing this behavior for $V=2$. On every clock cycle, the circuit can compute two output results.

Let us assume that $C$ can be clocked at a frequency that is $M$ times larger than the frequency of $clk_{0}$.
The multi-pumping optimization can be applied in two different ways, each affecting either the external or the internal, relative to the compute block being optimized, data paths of the design.
The first approach is where the widths of the internal data paths remain unchanged while the external widths are widened by the factor $M$. The internal computing part is driven by a clock signal $clk_1$, clocked $M$ times higher than $clk_0$. This scenario is depicted in waveform \two of \Cref{fig:waveform}, assuming $M=2$.
Data entering the multi-pumped domain must be converted from one wide vector of size $MV$ to $M$ narrow vectors of size $V$ --- and the inverse for leaving the multi-pumped domain (issuers and packers in \Cref{fig:waveform}). The resulting design obtains increased throughput by a factor $M$, at the same resource consumption as the original implementation. In the example, the circuit computes four output elements per clock cycle $clk_0$.

\begin{figure*}
    \centering
    \includegraphics[width=\textwidth]{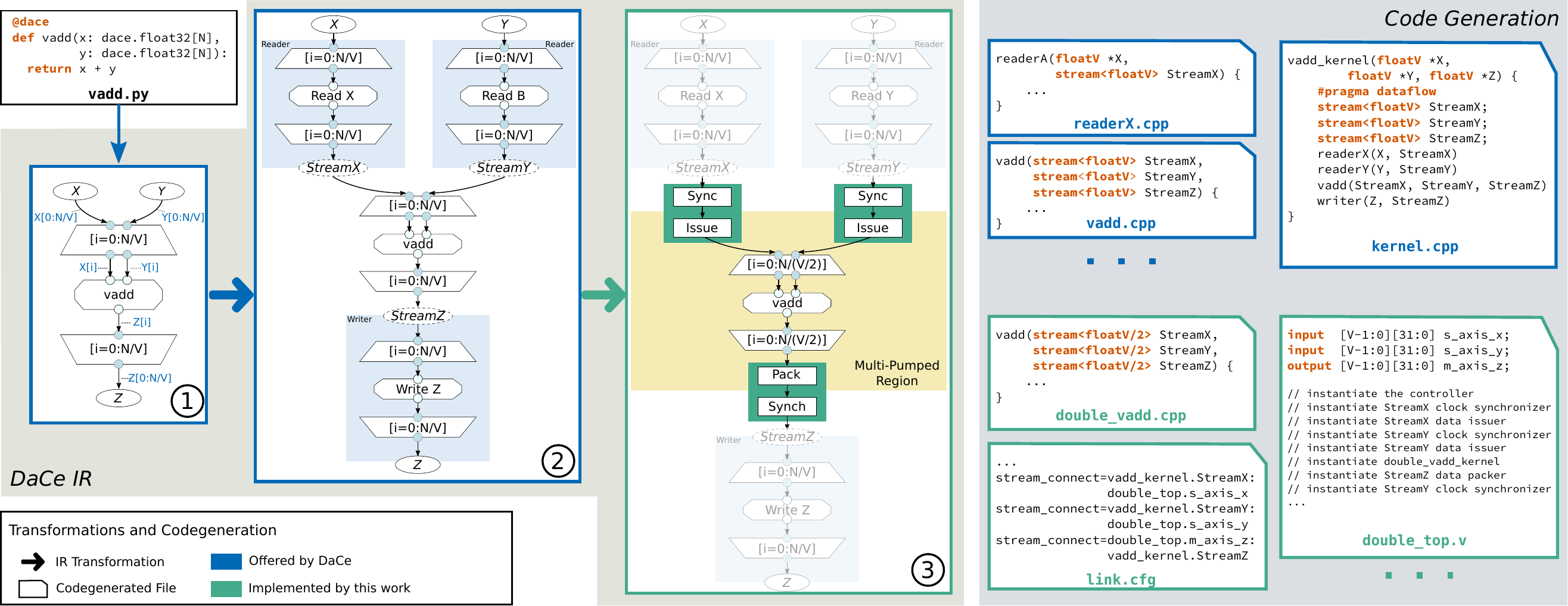}
    \caption{Workflow of automatic multi-pumping.}
    \label{fig:vadd-double-pump}
\end{figure*}

A second approach would be to divide the width of the paths internally in the compute blocks by the factor $M$, while the widths of the external paths remain unchanged (waveform \three in \Cref{fig:waveform}). The internal compute part runs according to $clk_1$, but we no longer need to use $V$ units of $C$ to keep up with the data rate since each of them can consume $M$ times the original number of data elements per tick of $clk_0$. The new implementation has the same throughput as the original one (two elements per clock cycle in the example), but at a reduced resource cost of $V/M$ units of $C$, compared to the previous cost of $V$ units.

Both approaches can be applied to vectorized computations, providing complementary results.
However, the former approach is more versatile as it can be conveniently applied to problems that cannot be easily vectorized, as the dependencies between operations are preserved.
FPGAs favor deep pipelines, and this approach further pushes the benefits of these deep pipelines: it enables feeding the circuit in a vectorized fashion without requiring the problem to be vectorized in a traditional way.
These gains, however, only apply if the design can meet the new increased clocking constraints, as the overall throughput otherwise decreases due to stalling.
We need to consider the \textit{effective clock rate} at which the circuit will operate, which can be defined as the minimum of the external clock rate ($clk_0$) and $1/M$ the internal clock rate ($clk_1)$. If the external components (readers/writers from memory) or internal ones (compute) cannot keep up, the other must stall.

In summary, the multi-pumping optimization does not come without a cost: synchronizing and multiplexing the data at the clock domain crossings introduces overhead. Whether this yields a net benefit depends on the size of the multi-pumped domain, the number and the size of data paths entering and leaving the domain, the overall constraining resource (i.e., introducing the optimization might cut DSP and BRAM usage at the cost of LUTs and registers), and the effective clock rate of the resulting design.

\section{Automating multi-pumping}
High-Level Synthesis (HLS) tools have improved the productivity in hardware programming, allowing the user to program in familiar languages such as C/C++ or OpenCL. However, vendor tools do not typically support multiple clocks~\cite{xilinx-ug902}, as required to implement multi-pumping, forcing programmers to utilize hand-tuned Register-Transfer Level (RTL) code. Instead, we use the Data-Centric (DaCe) parallel programming framework~\cite{dace} to productively expose automatic multi-pumping.

\subsection{Data-centric parallel programming}\label{sec:dace}
The DaCe framework revolves around a graph-based intermediate representation (IR) that captures all data movement of an input program in pure dataflow regions. This IR is produced from one of multiple frontends (such as Python) by symbolically analyzing the dataflow in the application, inherently exposing parallelism and memory access patterns.

Figure~\ref{fig:vadd-double-pump} details the workflow of automatic multi-pumping of a vector addition Python function using the DaCe IR. The program is first converted to the graph IR, representing data containers (random-access or streaming) and computation as nodes, and data movement between those components as edges. These edges describe data movement via symbolic expressions (blue edge labels in the figure). The IR natively maps computational components to FPGA modules, and represents parametric parallelism via Map scope nodes (trapezoids in the figure), which can be scheduled as multiple PEs or pipelined loops.

The IR is subsequently optimized by applying \textit{data-centric transformations}, either programmatically or interactively. Transformations are graph-rewriting rules that can check for feasibility and modify subgraphs in the IR. Vectorization, for example, is applied in Figure~\ref{fig:vadd-double-pump} (box \one): it changes the range of the parametric scope by dividing them by \texttt{V}, the applied vectorization factor; it converts the type of data containers to a vector data type;  and modifies the edges' addresses accordingly.
The DaCe IR can then be lowered to one of multiple backends, including Xilinx HLS C++ code (Figure~\ref{fig:vadd-double-pump}, right-hand side). DaCe automatically generates multiple files that contain the modules, govern the interface between modules, and an OpenCL-based host API for use in existing programs.

In this work, we exploit the explicit data movement given by the DaCe IR to detect program subgraphs where the multi-pumping optimization can be applied, and write a transformation that does so automatically. Following the transformation, we use DaCe to generate RTL modules, enabling multiple clock domains.

\subsection{Automatic transformation}\label{sec:auto-impl-steps}
Our automatic multi-pumping transformation applies to programs regardless of their computational contents, but rather by tracing and mutating their data movement properties. We summarize the steps of applying it in Figure \ref{fig:vadd-double-pump} and detail them below.

We identify where to apply the optimization by greedily taking the entire application in its DaCe IR form and finding the largest subgraph that can be \textit{streamed}, that is, when data dependencies between two components can be converted to queue-based access. We leverage DaCe's existing infrastructure for tracing module input and output index expressions. By performing an intersection check on each pair of connected modules, we can determine if pipelining the memory between two modules can be performed. By default, every such component in the given program will be streamed.

Once the computational modules are streamed, DaCe can automatically inject \textit{reader} and \textit{writer} modules externally to the computational part. The only change required is for it to work on these streams rather than accessing memory directly. For example, the vector addition example computing $z = x + y$ consists of a single loop, reading from $x$ and $y$, computing the result, and writing the result to $z$. To change it to a streaming implementation, the streaming transformation extracts the reads (writes) out of the computation by introducing other components that access $x$ and $y$ ($z$) in the same order as the computation, and push (pop) the values into streams. Box \two in Figure~\ref{fig:vadd-double-pump} shows the transformed graph. Now that the communication on the streams drives control flow, all the four components (two readers, compute, and writer) can run in parallel, allowing us to modify the rates of the interfaces to increase potential throughput.

To \textit{assess} the feasibility of whether the transformation can be applied to a streaming implementation, we build upon techniques used by compiler auto-vectorizers.  Therefore, the same conditions that apply to SIMD-capable code apply to temporally vectorizable (i.e., running the same operations on multiple data, predication instead of conditional control flow). Moreover, temporal vectorization is slightly more relaxed than the traditional vectorization --- as the instructions run in sequence (albeit faster), internal sequential dependencies across data are allowed. The only restriction is that the participating operations must not involve data-dependent external memory I/O based on previous operations.

We adapt the existing vectorization infrastructure in DaCe to perform the relaxed check and \textit{apply} the data-centric graph modification.
The only requirement we impose for the multi-pumped domain is that it follows a predefined set of interfaces that is instantiable from within RTL. In all our experiments, our computations are generated through HLS, using streaming AXI as the communication protocol. With that, we can modify the clock rate of our identified subgraph, retaining the non-multi-pumped components from the original design.

The remainder of the transformation performs global clock management. We inject ``plumbing'' modules that synchronize the data paths between the clock domains and distribute the data over the temporal dimension. We use the built-in Xilinx streaming AXI infrastructure IP cores~\cite{xilinx-axis-ip-library}, which implement three types of modules:

\begin{description}[leftmargin=2em]
    \item[Data synchronizers] These IP cores synchronize a data stream between two clock domains. We use them in both directions, so both going in and out of the multi-pumped domain. This is the first IP core in the chain, and the following ones run at the multiplied clock rate.
    \item[Data issuers] These IP cores divide a single transaction from a wide data stream into multiple transactions on a narrow data stream. We use these IP cores when moving data into the multi-pumped domain.
    \item[Data packers] These IP cores are the inverse of the issuers; they take multiple transactions from a narrow data stream and pack them onto a single transaction to a wide data stream. We use these IP cores when moving data out of the multi-pumped domain.
\end{description}

The three modules are customized based on the provided clock signal, and the widths of the data streams. We construct two clock domains: one domain for the readers and the writer and one for the computation core  (box \three in Figure~\ref{fig:vadd-double-pump}). For each stream handled by the readers, we insert an instance of a synchronizer and a data issuer in that order. For the stream handled by the writer, we insert an instance of a data packer and a synchronizer in that order.
The transformed graph is then passed to the code generation phase, which will emit the corresponding HLS/RTL code.

In summary, through data movement analysis, we are able to: \textit{a)} identify parallel subgraphs that do not access data beyond their own scope, except for external memory accesses; \textit{b)} extract these subgraphs into separate kernels that communicate through streams and can run in parallel; \textit{c)} automatically create different clock domains and inject hardware for handling domain crossings and temporally distribute and collect the crossing data, based on the external interfaces.

\subsection{HLS/RTL integration}
FPGA vendor tools typically follow the OpenCL programming model by having a host program and kernels which execute on the acceleration platform. The vendor tools can target HLS or RTL, but typically only RTL supports multiple clocks. To enable the most comprehensive set of optimizations from the vendor toolchain, we utilize HLS wherever possible in the computational part and only integrate RTL when necessary. There are two HLS/RTL integration approaches in the Xilinx toolchain: ``black-boxing'' (equivalent to inline assembly) or separate RTL kernels. We choose the latter since the former suffers from the same restrictions as regular HLS~\cite{xilinx-ug902}.

In this paper, we target Xilinx devices:
each HLS kernel is emitted by DaCe into a single C++ file. An RTL kernel is instead code-generated
into four different files:
\begin{enumerate}
    \vspace{-1mm}
    \item A SystemVerilog \emph{controller} for communicating with the host.
    \item A SystemVerilog \emph{computation core}.
    \item A Verilog \emph{top-level} file for instantiating both the controller and one or more instances of the computation core.
    \item A TCL \emph{script} for packaging the kernel.
    \vspace{-1mm}
\end{enumerate}
Multiple kernels can communicate directly with each other through streams, avoiding the need, and thus also the overhead, of the host program dictating the inter-kernel communication.
This is handled through a \emph{configuration} file describing, amongst the others, the kernels and how their interfaces connect together.
It is required that the RTL design follows a predefined interface in order to interact with other kernels, which is the streaming AXI protocol~\cite{streaming-axi}.

All the generated files are compiled using the vendor tools (\texttt{vitis\_hls} and \texttt{vivado}) and linked together based on the configuration file. Then, the produced device binary, consisting of the integrated HLS and RTL kernels, can be invoked by the host program.

We use the DaCe RTL backend as a wrapper around HLS code, allowing us to utilize RTL optimizations during the data organization of multiple clock domains while utilizing HLS for the computation core.
However, the code generation in its current form is not sufficient for implementing multi-pumping, requiring us to extend the FPGA backend of DaCe. In this regard, we have to:
\begin{description}[leftmargin=2em]
    \item[Integrate HLS cores in RTL.] DaCe RTL backend is limited to having only the computation cores written in RTL. To support cores written in HLS, we add the ability to include user IP libraries, as we can then emit the HLS cores as IP cores and instantiate these within RTL.
    \item[Instantiate multiple different cores in RTL.] In order to inject the "plumbing" IP cores, we need to instantiate the different controllers and the computation core in the top-level file.
    \item[Enable multiple clock and reset signals.] Given that we are implementing this in RTL, we are free to generate a clock generator and reset synchronizer based on our target clocks. However, this becomes a problem when having multiple kernels, as the clocking resources are quickly depleted, even for large boards. For designs requiring only two clock signals, we utilize the multiple clock signal capabilities of the Xilinx Vitis platform, where the shell can supply multiple clock and reset signals to the kernels, thus only consuming clocking resources once.
    \item[Automate the compilation flow.] We need to modify the compilation flow, as we now have more parts that need to be compiled individually before being linked together in the final device binary.
\end{description}

With the proposed extensions, DaCe, starting from the transformed IR, can automatically emit and compile all of the code necessary for multi-pumping, supporting multiple differently clocked RTL and HLS subdomains.

\subsection{When to apply the transformation}
Our primary strategy for the automatic application is taking the largest possible subgraph as the candidate for the transformation. This approach aims at minimizing the resource overhead introduced for synchronizing and multiplexing data while crossing multiple clock regions. On the other hand, this may increase the probability of routing congestion for large designs, which might lead to a reduced effective clock rate and reduced performance.
As mentioned in~\Cref{sec:dace}, the transformation can be applied both automatically and interactively, targeting smaller sub-computations. For the designs where the automatic application causes routing congestion, the user can guide the process towards different strategies that might counter these shortcomings.

The transformation can be applied multiple times to different parts of an entire design.
Thus, a viable strategy would be to identify bottlenecks of a design as candidates for the transformation, possibly utilizing the different approaches to the optimization for either throughput or resource consumption.
Applying the transformation to memory-bound problems would keep up with the same memory bandwidth while reducing the area. Conversely, applying it to compute-bound problems would increase the performance within the same amount of consumed resources.

For problems that cannot be easily traditionally vectorized due to dependent computations, we can apply this optimization without changing the internal computation, thus retaining the internal dependencies.
This allows us to feed data in a vectorized manner while packing computations temporally, increasing the overall throughput.

\section{Evaluation}

\sloppy We evaluate our approach on a Xilinx Alveo U280 accelerator. Kernels are built with Vitis \texttt{2020.2}, targeting the \texttt{xilinx\_u280\_xdma\_201920\_3} shell.
To remove external sources of congestion that are not under our direct control, unless otherwise specified we consider the following configuration:

\begin{description}
    \item[Single SLR] The U280 is a multi-chiplet FPGA,  where three Super Logic Regions (SLRs) are combined to form the overall chip. While this allows for larger chips, die crossing interconnects complicate the floor planning, lowering the maximum achievable frequency significantly. Therefore, for our evaluation we use a single SLR. The total amount of resource available in a single SLR is shown in Table~\ref{tab:resources}.
    \item[Direct access to HBM banks] The U280 is equipped with 32 HBM banks, all connected to \texttt{SLR0} ~\cite{alveo-u280}. Each bank is used exclusively to store a single container, so that we remove potential congestion that arises when multiple entities access the same memory bank.
\end{description}

\begin{table}[tb]
\begin{tabular}{ccccc}
\hline
\textbf{LUT Logic}    & \textbf{LUT Memory}   & \textbf{Registers}    & \textbf{BRAM}         & \textbf{DSPs}         \\ \hline
439\:K	&   205 \:K	    &   879\:K	&   672	    &   2,880\\
\bottomrule
\end{tabular}

\caption{Resources available for a single SLR (SLR0) of the Xilinx U280 FPGA.}
\label{tab:resources}
    \vspace{-5mm}
\end{table}

For each application, we report the resource utilization (percentage of the total amount) and the design frequencies, as declared by Vivado after the Place and Route stage. We derive the running time (reported in seconds) or performance (expressed in GOp/s) considering the actual execution time on the device. The term double-pumping refers to applying the multi-pumping optimization at a factor of two. For the double-pumped implementations, the effective clock rate becomes the minimum value of CL0 and half of CL1.  While our approach can also be applied with other pumping factors, for this evaluation we are limited by the maximum achievable frequency by Vivado (which is 650 MHz for the used version).

\subsection{Vector addition}\label{sec:experiments-vadd}

In this section, we report the results obtained by applying double-pumping to the vector addition example described in~\Cref{sec:auto-impl-steps}. Results are shown in Table~\ref{tab:vadd_results}.  As expected, the computational resource usage, i.e., DSPs, is halved at the cost of a marginal increase in LUT and Register consumption (less than 1\%). This is a good indication to the overhead of the synchronization, as the kernel itself is very simple.

\begin{table}

 \footnotesize
 \centering
 \resizebox{\columnwidth}{!}{
 \begin{tabular}{@{}lrrrrrr@{}}
    \toprule
                    & \multicolumn{2}{c}{\textbf{Vect. Width: 2}} & \multicolumn{2}{c}{\textbf{Vect. Width: 4}} & \multicolumn{2}{c}{\textbf{Vect. Width: 8}} \\
                    \cmidrule(l{1pt}r{1pt}){2-3} \cmidrule(l{1pt}r{1pt}){4-5} \cmidrule(l{1pt}r{1pt}){6-7}
                    & \textbf{O}          & \textbf{DP}          & \textbf{O}          &  \textbf{DP}          & \textbf{O}          &  \textbf{DP}          \\
    \midrule
    {Freq CL0 [MHz]} &   339.4   &   340   &  332.5   & 343.2   &    344.5   &   335.2 \\
    {Freq CL1 [MHz]} &   -   &   668.4   & -   & 651.4   &    -   &   643.9 \\
    \textbf{Time [s]}   &   0.1112  &   0.1111  &  0.0557  & 0.0557  &    0.0281  & 0.0280\\
    \midrule
    {LUT Logic [\%]} &   5.27    &   5.37    &  5.39    & 5.46    &    5.57    &   5.65 \\
    {LUT Memory [\%]}    &   2.27    &   2.26    &  2.34    & 2.33    &    2.48    &   2.47 \\
    {Registers [\%]}     &   6.74    &   6.95    &  6.86    & 7.16    &    7.05    &   7.57 \\
    \textbf{BRAM [\%]}  &   6.77    &   6.77    &  6.92    & 6.92    &    7.22    &   7.22 \\
    \textbf{DSP [\%]}   &   0.14    &   0.07    &  0.28    & 0.14    &    0.56    &   0.28 \\
    \bottomrule
    \end{tabular}
    }
    \caption{Results for vector addition, Original (O) and Double Pumped (DP) versions.}
    \label{tab:vadd_results}

    \vspace{-5mm}
 \end{table}

\subsection{Matrix Multiplication}

Matrix Multiplication is a widely studied problem in the FPGA community. To evaluate the benefits of our multi-pumping transformation, we consider the state-of-the-art Communication Avoiding Matrix Multiplication proposed by de Fine Licht et al.~\cite{ca-mmm}.
We implement this approach in DaCe, using a one-dimensional systolic array, targeting 32-bit floating point numbers. The systolic array contains a configurable number of vectorized Processing Elements (PE).  Every PE is only connected to the previous and the next, and passes data along the chain from the head towards the tail. The systolic array feeders, PEs, and drainers implement the proposed I/O optimal strategy.
The first two columns of \Cref{tab:gemm_results} compare the DaCe implementation against the hand-written Xilinx HLS implementation provided by the authors and which is publicly available at~\cite{ca-mm-code}. We report the best performing configuration for a single SLR (notably, we fix the PE vectorization width to 16 to retain maximum performance and scaled the number of PEs and tiles sizes to fill up a single SLR). The table shows both implementations perform on par, demonstrating the quality of the baseline code generated by DaCe.

We apply double pumping targeting the entire systolic array, leaving only the readers/writers to/from global memory in the slowly clocked domain, with the computations in the higher clocked domain.
We expect that by applying the optimization to a larger subdomain, we will see a reduction of resources on more than one type of resource compared to the vector addition example. Results are shown in~\Cref{tab:gemm_results}.

\begin{table}[t]
 \small
 \centering
 \begin{tabular}{@{}lrrr|rr@{}}
    \toprule
                    &\multicolumn{3}{c}{\textbf{32 PEs}} &
                    \textbf{48 PEs} & \textbf{64 PEs} \\
                    \cmidrule(l{1pt}r{1pt}){2-4} \cmidrule(l{1pt}r{1pt}){5-5} \cmidrule(l{1pt}r{1pt}){6-6}
                    & \textbf{CA \cite{ca-mmm}}   &   \textbf{O}          & \textbf{DP}         & \textbf{DP}          &  \textbf{DP} \\
    \midrule
    {Freq CL0 [MHz]} 	&	250 	&	268 	&	261.4 	&	269.9 	&	252.9\\
    {Freq CL1 [MHz]} 	&	- 	&	- 	&	452.8 	&	398.2 	&	322.5\\
    \textbf{Perf [GOp/s]} 	&	253.2 	&	256.1 	&	219.1 	&	260.8 	&	293.8	\\
    \midrule
    {LUT Logic [\%]} 	&	43.9 	&	44.8 	&	32.1 	&	41.3 	&	53.7\\
    {LUT Memory [\%]} 	&	6.9 	&	13 	&	10.1 	&	14.8 	&	17.4\\
    {Registers [\%]} 	&	44.5 	&	44.3 	&	36.6 	&	45.9 	&	60.1\\
    \textbf{BRAM [\%]} 	&	81.4 	&	80.3 	&	47 	&	63.6 	&	82.7\\
    \textbf{DSP [\%]} 	&	88.9 	&	90 	&	45.6 	&	67.9 	&	90\\
    \midrule
    \textbf{MOp/s per DSP}    &	98.9    &	98.8    &   167.0   &	133.5   &	113.3\\
    \bottomrule
    \end{tabular}
    \caption{Results for Matrix-Matrix Multiplication: Communication Avoiding~\cite{ca-mmm} (CA), DaCe Original (O) and Double Pumped (DP) versions.}
    \label{tab:gemm_results}
    \vspace{-10mm}
 \end{table}

Considering the same number of PEs, the double pumped versions consume fewer resources (down to ${\approx}50\%$ and ${\approx}58\%$ of DSPs and BRAM, respectively). As for the two clock domains, the double-pumped version runs at an effective clock rate that is lower than the original one, resulting in slightly lower performance.
However, the reduced resource utilization of the double-pumped version allows us to scale the design further, increasing the number of PEs by up to $64$ (the last two columns in the table). For larger systolic arrays, the increased congestion can impair the achieved maximum frequency, resulting in the version with 64 PEs performing only slightly better than the version with 48 PEs. Using multi-pumping, a speedup of $15\%$ is achieved automatically over the HLS hand-written and DaCe versions.

We consider the performance attained by a DSP (measured in MOp/s per DSP) as a metric to evaluate the benefits of multi-pumping on this resource type critical for compute-intensive applications. As can be noted, with multi-pumping DSPs are exploited better than in the original version. Highly-congested design can bring reduced returns due to a lower effective frequency, but we still increase the DSP performance with respect to the original design.

To verify the scalability of our approach to full chip designs, we replicate the version with 64 PEs across three SLRs, each performing an independent computation. Our results show that we reach around 477.3 GOp/s, which is 25\% scaling efficiency with respect to the single SLR version, due to the clock rate being impaired by SLRs crossing. To counter this, it would be interesting to apply the SLR crossing optimizations offered by AutoBridge~\cite{21-autobridge}, as this would reduce the overhead introduced by crossing SLRs. Regardless, the multi-pumping optimization results in a a 18\% speedup compared to the hand-written HLS implementation, able to achieve 401.4 GOp/s across three SLRs.

\subsection{StencilFlow}
StencilFlow~\cite{stencilflow} is a state-of-the-art domain-specific framework built on top of DaCe, designed to map directed acyclic graphs of stencil computations to FPGA systems.
We  benchmark the Jacobi 3D and Diffusion 3D iterative stencil programs, with  32-bit  floating  point  types. Due to varying arithmetic intensity, we apply 8-way spatial vectorization to Jacobi (which exhibits lower arithmetic intensity) and 4-way for Diffusion (higher intensity). Similarly to~\cite{stencilflow}, we chain together long linear sequences of stencils executed on a large input domain
of $2^{16}\times32\times32$ points. We increase the number of chained stencil computations until the SLR is fully utilized. Each stage in the chain is isolated in its own domain, requiring synchronization steps in between each stage. This also means that each step becomes its own independent kernel.
Regardless of having multiple synchronization steps, we still expect to see reduced resources, but not at as much as matrix multiplication.  Results are reported in Tables~\ref{tab:jacobi_results} and \ref{tab:diffusion_results}.

\begin{table}[htb]

 \footnotesize
 \centering
 \resizebox{\columnwidth}{!}{
 \begin{tabular}{@{}lrrrr|rr@{}}
    \toprule
                    & \multicolumn{2}{c}{\textbf{S=8}} & \multicolumn{2}{c}{\textbf{S=16}} &
                    \textbf{S=24} & \textbf{S=40}\\
                    \cmidrule(l{1pt}r{1pt}){2-3} \cmidrule(l{1pt}r{1pt}){4-5} \cmidrule(l{1pt}r{1pt}){6-6} \cmidrule(l{1pt}r{1pt}){7-7}
                    & \textbf{O}          & \textbf{DP}          & \textbf{O}          &  \textbf{DP}          & \textbf{O}          &  \textbf{DP}          \\
    \midrule
    {Freq CL0 [MHz]} &	307.6	&	322.4	&	304.2	&	331.5	&	305	&	258\\
    {Freq CL1 [MHz]} &	-	&	510.4	&	-	&	478	&	-	&	460.8\\
    \textbf{Perf [GOp/s]}	&	101.4	&	96.9	&	202.5	&	180.7	&	245.3	&	414.8\\
    \midrule
    {LUT Logic [\%]} &	20.25	&	14.2	&	36.15	&	23.37	&	42.17	&	47.78\\
    {LUT Memory [\%]}    &6.21	&	6.89	&	10.58	&	12.01	&	12.71	&	26.1\\
    {Registers [\%]}     & 22.48	&	19.14	&	39.21	&	32.5	&	49.2	&	64.5\\
    \textbf{BRAM [\%]}  & 15.33	&	10.57	&	24.85	&	15.33	&	30.11	&	23.41\\
    \textbf{DSP [\%]}   &	28.89	&	14.44	&	57.78	&	28.89	&	72.22	&	72.22\\
    \midrule
    \textbf{MOp/s per DSP}   &	121.9	&   232.8	&   121.7	&   217.1	&   117.9	&   199.0\\
    \bottomrule
    \end{tabular}
    }
    \caption{Results for Jacobi 3D Stencil, Original (O) and Double Pumped (DP) versions, considering different number of chained stencils (S).}
    \label{tab:jacobi_results}
    \vspace{-5mm}
\end{table}
\begin{table}[htb]

 \footnotesize
 \centering
 \resizebox{\columnwidth}{!}{
 \begin{tabular}{@{}lrrrr|rr@{}}
    \toprule
                    & \multicolumn{2}{c}{\textbf{S=8}} & \multicolumn{2}{c}{\textbf{S=16}} &
                    \textbf{S=20} & \textbf{S=40}\\
                    \cmidrule(l{1pt}r{1pt}){2-3} \cmidrule(l{1pt}r{1pt}){4-5} \cmidrule(l{1pt}r{1pt}){6-6} \cmidrule(l{1pt}r{1pt}){7-7}
                    & \textbf{O}          & \textbf{DP}          & \textbf{O}          &  \textbf{DP}          & \textbf{O}          &  \textbf{DP}          \\
    \midrule
    {Freq CL0 [MHz]} &	309.1	&	329.4	&	311.4	&	333.1	&	305	&	255.2 \\
    {Freq CL1 [MHz]} &	-	&	537.3	&	-	&	490.4	&	-	&	462.9 \\
    \textbf{Perf [GOp/s]}	&	110.4	&   102.8	&	220.6	&	202.6	&	275.7	&	460.3 \\
    \midrule
    {LUT Logic [\%]} &	16.55	&	12.08	&	28.52	&	19.42	&	34.57	&	40.66 \\
    {LUT Memory [\%]} &	4.85	&	5.27	&	7.91	&	8.8	&	9.44	&	19.38 \\
    {Registers [\%]}     &	18.25	&	15.88	&	30.96	&	25.94	&	37.27	&	56.12 \\
    \textbf{BRAM [\%]}  &	10.57	&	8.18	&	15.33	&	10.57	&	17.71	&	17.71 \\
    \textbf{DSP [\%]}   &	31.67	&	16.67	&	63.33	&	33.33	&	79.17	&	83.33 \\
    \midrule
    \textbf{MOp/s per DSP}    & 121.0	&   214.2	&   121.0	&   211.1	&   120.9	&   191.8\\
    \bottomrule
    \end{tabular}
    }
    \caption{Results for Diffusion 3D Stencil, Original (O) and Double Pumped (DP) versions, considering different number of chained stencils (S).}
    \label{tab:diffusion_results}
    \vspace{-5mm}
\end{table}

As with Matrix Multiplication, double-pumping stencils allows for a transparent reduction in resource consumption, at the cost of slightly reduced performance. For all of the double-pumped versions, the performance per DSP has increased by more than  $50 \%$ compared to the original implementation. For both stencils, we have been able to exploit the additional available resources to scale designs even further, allowing for increased performance at 69\% and 66\% respectively.

\subsection{Floyd-Warshall}
The temporal vectorization approach does not make any assumptions about the computation in the multi-pumped domain, and therefore it can also be applied to programs that are not easily vectorizable with traditional (dependent or independent) vectorization.
We consider the Floyd-Warshall algorithm, which computes the shortest paths between each pair of nodes in a graph. Deriving a vectorized implementation for such algorithm requires dealing with the iteration dependencies that characterize its nested loop. While this can be done by an experienced programmer, we can use the multi-pumping transformation to gain performance without modifying the implementation. As discussed in Section~\ref{sec:exploiting-clock-domains}, the optimization does not reduce resource consumption, but increases performance. The results are shown in~\Cref{tab:floyd}.

\begin{table}[htb]

 \small
 \centering
 \begin{tabular}{@{}lrr@{}}
    \toprule
                    & \textbf{O} & \textbf{DP}\\
    \midrule
    {Freq CL0 [MHz]} &	527.9	&	520.2	\\
    {Freq CL1 [MHz]} &	-	&	674.7	 \\
    \textbf{Time [s]}	&	5.02	&   3.36	 \\
    \midrule
    {LUT Logic [\%]} &	5.35	&	5.45	\\
    {LUT Memory [\%]} &	2.22	&	2.29	\\
    {Registers [\%]}     &	6.38	&	6.67	\\
    \textbf{BRAM [\%]}  &	34	&	32	\\
    \textbf{DSP [\%]}   &	0.14	&	0.21	\\
    \bottomrule
    \end{tabular}
    \caption{Results for Floyd-Warshall algorithm (graph with 500 nodes),  Original (O) and Double Pumped (DP) versions.}
    \label{tab:floyd}
    \vspace{-5mm}
\end{table}
The results in the table match our expectations:  with similar resource consumption, we are able to increase our performance by $50\%$. The performance gain cannot be higher, as we are limited by the maximum frequency we can request to Vitis (650 MHz for the version used in this evaluation).

\subsection{Discussion}

\begin{figure}[b]
    \centering
    \includegraphics[width=\columnwidth]{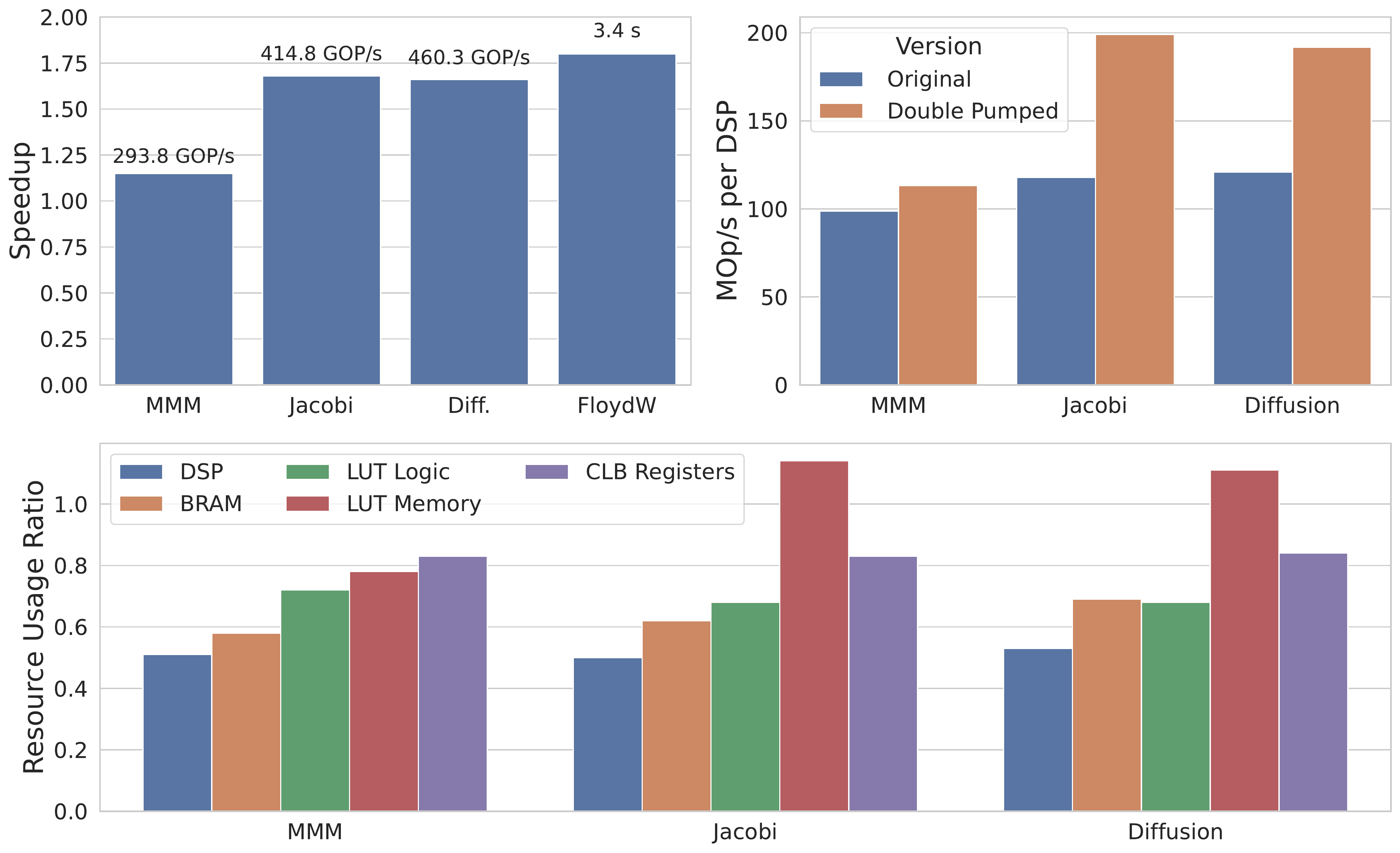}
    \caption{Performance and resource-saving overview. First row: the speedup and the DSP efficiency for the best performing double-pumped and original versions. Labels report absolute performance figures. Second row: ratio of resources used by double pumped version over the original one, considering the same application configuration: 32 PEs for Matrix Multiplication (MMM), 16 stages for stencils.}
    \label{fig:results_overview}
\end{figure}

\figurename~\ref{fig:results_overview} summarizes the performance and resource-saving benefits of our proposed approach.
Generally, applying multi-pumping reduces the amount of consumed critical resources up to $50\%$, as shown in \Cref{fig:results_overview}, second row. For replicated designs, this reduction allows for further replication, which translates into an increase in overall performance, as seen in the matrix multiplication and stencil experiments. For designs that are not easily vectorized in the traditional sense, such as the Floyd-Warshall example, this automatic optimization allows vectorization across the temporal dimension.
This should enable even higher frequencies for deeply pipelined designs.
Furthermore, looking at the tables of the different samples, we see only a marginal resource consumption of the non-critical resources - even for the stencil examples which utilize multiple kernels that need to communicate.

By applying the multi-pumping optimization, we obtain a better exploitation of the critical resources, as the MOp/s per DSP bar plot depicts.
Generally, the reduced amount of consumed resources can relax the degree of congestion, pushing the clock rates of the internal clock domain higher. This is prominent in all of the experiments, as the CL1 of the double-pumped versions are higher than the CL0 of the original version. This, however, does not automatically translate into increased performance, as the effective clock rate becomes a limitation, which our experiments show primarily because the original counterparts already reach high clock rates.
While we have not directly power consumption, we do not see any significant increase when using the same configurations, indicating light power consumption overhead. For the designs that are replicated to utilize the full SLR, we see about 10\% increase. However, we would have to perform deeper analysis to justify this claim.

While DaCe enables and eases the application of this optimization, we should note that it is not limited to DaCe and should be applicable in other frameworks, such as HLS, but \textit{only if memory accesses can be traced across modules}, which current compilation pipelines inhibit. Moreover, in its current implementation, the transformation can be automatically applied to either HLS or RTL cores through DaCe, but could fundamentally be applied to a larger set of computation cores. As long as the core uses streaming AXI interfaces, and can be instantiated within RTL, the optimization can be applied. This work will be integrated into the core of DaCe, allowing other DaCe applications to use this optimization technique.

\section{Related Work}
Multi-pumping has been traditionally used as a resource-sharing technique for FPGA designs: hardware components, clocked at multiples of the surrounding logic, are shared in a time-multiplexed manner by mapping multiple operations per clock cycle.
Ronak et al.~\cite{16-multidsp} apply multi-pumping on DSP blocks. They propose an intermediate graph representation (DSP DataFlow Graphs),  where nodes represent operations that can be mapped on multi-pumped  DSPs. They save up to 48\% of DSPs but use up to 74\% additional LUTs.
Other approaches ~\cite{12-mipscache, 13-multireg} apply multi-pumping to BRAM to allow multiple reads/writes per clock cycle.
DeepPump~\cite{17-deeppump} generates hardware designs of Convolutional Neural Networks with multi-pumped computation units, resulting in improved throughput, but increased power consumption, and double the latency.
FTDL~\cite{20-ftdl} is an FPGA overlay framework for deep learning applications, which exploits double-pumping for logic blocks, DSPs, and BRAM. It allows for higher hardware efficiency and improved performance over competitors.
Anderson et al.~\cite{21-cgra} propose a Coarse-Grained Reconfigurable Architecture, allowing for these coarse blocks to use low-level optimizations, such as multi-pumping. Their implementation does not improve performance but does reduce area.
While these approaches increase performance or improve area utilization, they all require using highly- and manually-tuned RTL implementation to exploit the multi-pumping optimization.

A few solutions lift multi-pumping at higher abstraction levels, including it as an HLS directive. LegUp HLS~\cite{13-legup}, allows for DSP and BRAM double-pumping, by constraining the maximum allowed amount of resources.
The Intel OpenCL compiler offers BRAM double-pumping through the \texttt{doublepump} attribute~\cite{intel-double}. It does not increase performance but increases the number of accesses per clock cycle that can be issued to a BRAM block.
These solutions offer multi-pumping to HLS programmers, but it can still be applied only at very fine granularity (single DSPs, BRAM). Furthermore, pragmas or compiler hints leave to programmers the burden of deciding where and how to apply such optimization. In contrast, we show how multi-pumping can be applied to entire computational subdomains, and how to exploit data movement to automatically guide this process.

\section{Conclusion}
In this work, we show how data movement analysis can be used to automatically apply the multi-pumping optimization, which schedules critical computational subdomains to a higher clock, while widening the data path of the surrounding logic. We demonstrated that automatic multi-pumping reduces resource usage by up to 50\% in the higher clocked subdomain without sacrificing performance, or even increase overall performance in state-of-the-art designs by using the saved resources to increase parallelism in the design. We equate our transformation to a form of \textit{temporal} vectorization, where the computational itself is ``vectorized'' across multiple clock cycles fed by wider data paths across the chip. Temporal vectorization has a wider applicability with respect to traditional vectorization, and, by automating this process, engineers gain access to potentially free resource savings or a performance boost in designs that can be multi-pumped.

\section*{Acknowledgements}
\begin{wrapfigure}{l}{.3\linewidth}
    \centering
    \includegraphics[width=1.2\linewidth]{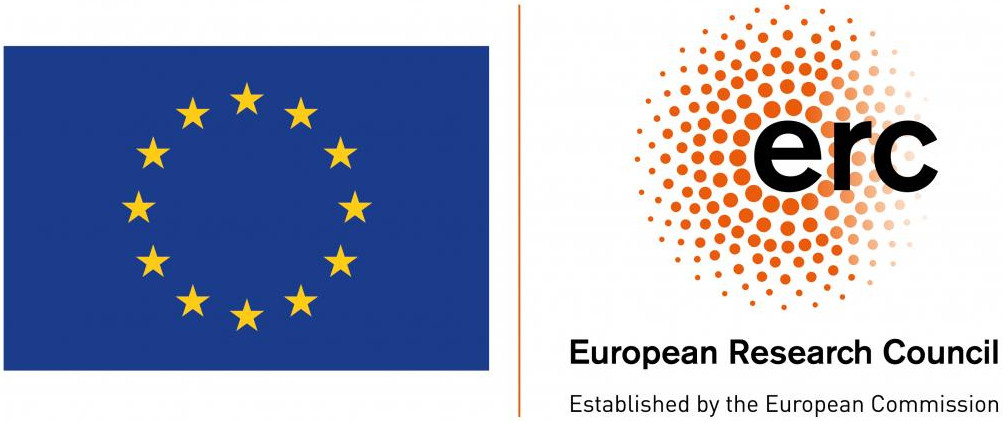}
    \vspace{-10pt}
\end{wrapfigure}
This project received funding from the European Research Council (ERC) grant PSAP, grant agreement No. 101002047, and the European Union’s Horizon Europe programme DEEP-SEA, grant agreement No. 955606.

\noindent
C.J. was funded by the Innovation Fund Denmark (IFD) under File No. 8057-00012B, the IFD Grand Solutions project ``Adaptive X-ray InSpection''.
\\
T.B.N. is supported by the Swiss National Science Foundation (Ambizione Project \#185778).
\\
We acknowledge the Xilinx University Program for providing
access to the AMD Xilinx Heterogeneous Accelerated Compute Cluster (HACC) at ETH Zurich.

\bibliographystyle{acm}
\bibliography{temporal_vectorization.bib}

\end{document}